\documentclass[11pt]{article}
\usepackage{amssymb,amsmath,amsfonts}
\usepackage{graphicx}
\usepackage{graphics}
\usepackage{epsfig}

\makeatletter

\begin{document}

\title{Fermionic condensate and the vacuum energy-momentum tensor for planar
fermions in homogeneous electric and magnetic fields}
\author{V. V. Parazian\thanks{%
E-mail: vparazian@gmail.com} \\
%EndAName
\textit{Institute of Applied Problems of Physics NAS RA,}\\
\textit{25 Nersessian Street, 0014 Yerevan, Armenia}}
\maketitle

\begin{abstract}
We consider a massive fermionic quantum field localized on a plane in
external constant and homogeneous electric and magnetic fields. The magnetic
field is perpendicular to the plane and the electric field is parallel. The
complete set of solutions to the Dirac equation is presented. As important
physical characteristics of the vacuum state, the fermion condensate and the
expectation value of the energy-momentum tensor are investigated. The
renormalization is performed using the Hurwitz function. The results are
compared with those previously studied in the case of zero electric field.
We discuss the behavior of the vacuum expectation values in different
regions for the values of the problem parameters. Applications of the
results include the electronic subsystem of graphene sheet described by the
Dirac model in the long-wavelength approximation.
\end{abstract}

\bigskip


\bigskip

\section{Introduction}

The planar fermions appear in describing the properties of a number of
condensed matter systems, such as graphene-type materials, topological
insulators, etc. Related (2+1)-dimensional quantum field theories also serve
as simplified models in elementary particle physics. Among the interesting
directions of the corresponding investigations is the study of the behavior
of planar fermions in external classical fields. The latter may generate a
number of interesting physical effects. In condensed matter systems,
described by the effective theory, the ground state corresponds to the
vacuum state in the related quantum field theory. The properties of the
zero-point fluctuations of the fields in the vacuum state depend on the
external fields and therefore the physical characteristics of that state
also will depend on them. For fermionic fields, among the important local
characteristics are the fermionic condensate and the vacuum average of the
energy-momentum tensor. In particular, the fermion condensate is the order
parameter describing the phase transitions. In the present paper, we
consider a massive fermionic quantum field localized on a plane in external
constant and homogeneous electric and magnetic fields.

The quantization of fields requires the knowledge of a complete set of the
mode functions being the solutions of the field equation. The solutions of
the Dirac equation in external electromagnetic fields have been considered
in different cases: (a) Coulomb potential, (b) constant magnetic field, (c)
constant electric field, (d) plane wave field, (e) plane wave field with
constant magnetic field along the direction of propagation, (f) four cases
in which electromagnetic potentials are assumed to have a special functional
dependence on the coordinates that lead to the equations being solved, (g)
constant electric and magnetic fields orthogonal to each other (see \cite%
{PAMDirac},\cite{Lukose} and references therein). The properties of a
massive fermionic quantum field localized on a plane in external constant
and uniform magnetic fields have been extensively investigated in literature
(see, e.g., \cite{Shovkovy} and references therein). For theoretical
investigation, the magnetic fields are a good tool and can be used for many
applications, for example, heavy ion collisions, quasi-relativistic
condensed matter systems like graphene, the Early Universe, and neutron
stars. The magnetic fields enhance (or "catalyze") spin-zero
fermion-antifermion condensates, associated with the breaking of global
symmetries (e.g., such as the chiral symmetry in particle physics and the
spin-valley symmetry in graphene) and lead to a dynamical generation of
masses (energy gaps) in the (quasi-)particle spectra. It was found that a
constant magnetic field stabilizes the chirally broken vacuum state.

The influence of a constant magnetic field on the dynamical flavor symmetry
breaking in 2 + 1 dimensions has been studied in \cite{Gusynin}. It is shown
that a fermion dynamical mass is generated even at the weakest attractive
interaction between fermions. This effect is considered in the
Nambu-Jona-Lasinio model in a magnetic field. The authors derived the
low-energy effective action and established the thermodynamic properties of
the model.

The vacuum current density and the condensate induced by external static
magnetic fields in (2+1)-dimensions were investigated in \cite{Raya}. The
authors discuss an exponentially decaying magnetic field along one of the
cartesian coordinates at the perturbative level. Non-perturbatively, by
solving the Schwinger-Dyson equation in the rainbow-ladder approximation
they get the fermion propagator in the presence of a uniform magnetic field.
In agreement with early expectations, in the large flux limit both these
quantities, either perturbative (inhomogeneous) and non-perturbative
(homogeneous), are proportional to the external field.

The competing effects of strains and magnetic fields in single-layer
graphene are discussed in \cite{Sanchez} to explore their influence on
different phenomena in quantum field theory, like induced charge density,
magnetic catalysis, dynamical mass generation, and magnetization. Among the
effects of the interaction between strains and magnetic fields, chiral
symmetry breaking, parity, and time-reversal symmetry breaking are
considered. The dynamical generation of a Haldane mass term and induced
valley polarization are also discussed. (2 + 1)-dimensional QED combined
with Dirac fermions both at zero and finite temperature has been considered
in \cite{Cea}. The effective Lagrangian and mass operator for planar charged
fermions in a constant external homogeneous magnetic field in the one-loop
approximation of the (2+1)-dimensional quantum electrodynamics are discussed
in \cite{Khalilov}. It has been established that fermionic masses might be
generated dynamically in an external magnetic field if the charged fermion
has a small bare mass.

In \cite{Lenz} has been shown that the growth of the chiral condensate with
the magnetic field in the (2 + 1)-dimensional Gross-Neveu model is preserved
beyond the mean-field limit for temperatures below the chiral phase
transition, and the critical temperature increases with increasing magnetic
field. In 2 + 1 dimensions a constant magnetic field is a strong catalyst of
dynamical flavor symmetry breaking, that is responsible for generating a
fermion dynamical mass even at the weakest attractive interplay between
fermions. The effect is illustrated in the Nambu-Jona-Lasinio model in a
magnetic field \cite{Abhshek}.

Several mechanisms for the formation of the FC have been studied in the
literature. They involve diverse kinds of interplays of fermionic fields, in
particular, the Nambu--Jona-Lasinio-type models with self-interacting
fields. In some models, the FC is linked to the gauge field condensate
(gluon condensate in quantum chromodynamics). The fermionic condensate is an
important characteristic in models with interacting fermions. It arises as
an order parameter in the corresponding phase transition. The change in the
sign of the condensate is the indication on a phase transition in the system
(see, e.g., \cite{Feng}).

The present paper is organized as follows. In the next section, we present
the complete set of fermionic modes in external static and homogeneous
electric and magnetic fields. The magnetic field is perpendicular to the
plane of fermion location and the electric field lies in that plane. By
using those modes, the fermion condensate is investigated in Section \ref%
{sec:FC}. The renormalization is based on the related Hurwitz function. The
respective investigation for the vacuum expectation value of the
energy-momentum tensor is presented in Section \ref{sec:EMT}. Section \ref%
{sec:Conc} summarizes the main results of the paper.

\section{Mode functions for planar fermions in homogeneous electric and
magnetic fields}

\label{sec:Modes}

In the presence of an external electromagnetic field with the vector
potential $A_{\mu }$ the Dirac equation for a quantum fermion field $\Psi
(x) $ is presented as
\begin{equation}
\left( i\gamma ^{\mu }D_{\mu }-sm\right) \Psi (x)=0,  \label{Dirac}
\end{equation}%
where $D_{\mu }=\partial _{\mu }+ieA_{\mu }$, (in units $\hbar =c=1$) is the
gauge extended covariant derivative and $e$ is the charge of the field
quanta. The Dirac matrices $\gamma ^{\mu }$ obey the Clifford algebra $%
\{\gamma ^{\mu },\gamma ^{\nu }\}=2g^{\mu \nu }$. As \ the background
geometry, we consider (2+1)-dimensional flat spacetime with the metric
tensor $g_{\mu \nu }=\mathrm{diag}(1,-1,-1)$. It is well known that in the
odd number of spacetime dimensions the Clifford algebra has two inequivalent
irreducible representations (with $2\times 2$ Dirac matrices in (2+1)
dimensions). We shall discuss the case of a fermionic field realizing the
irreducible representation of the Clifford algebra and, hence, $\Psi (x)$ is
a two-component spinor. The parameter $s$ in Eq. (\ref{Dirac}), with the
values $s=+1$ and $s=-1$, corresponds to two different representations. With
these representations, the mass term violates both the parity ($P$-) and
time-reversal ($T$-) invariances. In the long wavelength description of the
graphene, $s$ labels two Dirac cones corresponding to $\mathbf{K}_{+}$ and $%
\mathbf{K}_{-}$ valleys of the hexagonal lattice.

For the external electromagnetic field, we consider a configuration
described in Cartesian coordinates $\left( x^{0},x^{1},x^{2}\right) =\left(
t,x,y\right) $ by the vector potential $A_{\mu }=\left(
A_{0},A_{1},A_{2}\right) =\left( -Ex,0,Bx\right) $, where $E,$ $B=\mathrm{%
const}$. This corresponds to a constant electric field directed along the $x$%
-axis and a magnetic field perpendicular to the plane $(x,y)$. For the Dirac
matrices, we will take the representation
\begin{equation}
\gamma ^{0}=\sigma _{3},\;\gamma ^{1}=i\sigma _{2}=\left(
\begin{array}{cc}
0 & 1 \\
-1 & 0%
\end{array}%
\right) ,\;\gamma ^{2}=-i\sigma _{1}=\left(
\begin{array}{cc}
0 & -i \\
-i & 0%
\end{array}%
\right) ,  \label{gamma matrices}
\end{equation}%
where $\sigma _{1}$, $\sigma _{2}$, $\sigma _{3}$ are the Pauli matrices.
For the covariant components we have $\gamma _{\mu }=g_{\mu \nu }\gamma
^{\nu }=\left\{ \gamma ^{0},-\gamma ^{1},-\gamma ^{2}\right\} $. Assuming
that $|\beta |<1$, with $\beta =E/B$, we can boost to the reference frame
where the electric field vanishes and use the fermionic modes in the problem
with a homogeneous magnetic field. The corresponding coordinates will be
denoted by $\left( \tilde{x}^{0},\tilde{x}^{1},\tilde{x}^{2}\right) $. The
inverse boost will give the mode functions in the initial reference frame.
This procedure has been used in \cite{Lukose} to investigate the
corresponding energy spectrum.

We apply a Lorentz boost in the $y$ direction (perpendicular to the electric
field),
\begin{equation}
\left(
\begin{array}{c}
\tilde{x}^{0} \\
\tilde{x}^{2}%
\end{array}%
\right) =\left(
\begin{array}{cc}
\cosh \theta & \sinh \theta \\
\sinh \theta & \cosh \theta%
\end{array}%
\right) \left(
\begin{array}{c}
x^{0} \\
x^{2}%
\end{array}%
\right) ,  \label{Lorentzboostcoor}
\end{equation}%
and \ $\tilde{x}^{1}=x^{1}$. For the wave function we have the
transformation $\tilde{\Psi}\left( \tilde{x}^{\mu }\right) =\left( \tilde{%
\Psi}_{\uparrow }\left( \tilde{x}^{\mu }\right) ,\tilde{\Psi}_{\downarrow
}\left( \tilde{x}^{\mu }\right) \right) ^{T}=\exp \left( \theta \sigma
_{2}/2\right) \Psi \left( x^{\mu }\right) $, where $T$ stands for
transponation.

Applying the above transformation and choosing $\tanh \theta =E/B=\beta $,
we can rewrite the Dirac equation in the form
\begin{equation}
\left( \gamma ^{0}\tilde{\partial}_{0}+\gamma ^{1}\tilde{\partial}%
_{1}+\gamma ^{2}\tilde{\partial}_{2}+ie\gamma ^{2}\tilde{B}\tilde{x}%
^{1}+ism\right) \tilde{\Psi}(\tilde{x}^{\mu })=0,  \label{labDirac}
\end{equation}%
where $\tilde{B}=B\sqrt{1-\beta ^{2}}$ is the magnetic field in the boosted
frame. In the discussion below we will assume that $e\tilde{B}>0$. We
present the solution of equation (\ref{labDirac}) in the form $\tilde{\Psi}(%
\tilde{x}^{\mu })=e^{i\left( \tilde{p}^{2}\tilde{x}^{2}-\tilde{p}^{0}\tilde{x%
}^{0}\right) }\left( \phi _{\uparrow }\left( \tilde{x}^{1}\right) ,\phi
_{\downarrow }\left( \tilde{x}^{1}\right) \right) ^{T}$, where $-\infty <%
\tilde{p}^{2}<+\infty $. For the separate components from (\ref{labDirac})
we obtain%
\begin{eqnarray}
\tilde{\partial}_{1}\phi _{\downarrow }\left( \tilde{x}^{1}\right) +\left(
\tilde{p}^{2}+e\tilde{B}\tilde{x}^{1}\right) \phi _{\downarrow }\left(
\tilde{x}^{1}\right) -i\left( \tilde{p}^{0}-sm\right) \phi _{\uparrow
}\left( \tilde{x}^{1}\right)  &=&0,  \notag \\
\tilde{\partial}_{1}\phi _{\uparrow }\left( \tilde{x}^{1}\right) -\left(
\tilde{p}^{2}+e\tilde{B}\tilde{x}^{1}\right) \phi _{\uparrow }\left( \tilde{x%
}^{1}\right) -i\left( \tilde{p}^{0}+sm\right) \phi _{\downarrow }\left(
\tilde{x}^{1}\right)  &=&0.  \label{equations phys}
\end{eqnarray}%
Assuming $\tilde{p}^{2}\neq \pm sm$, one gets the equation for the upper
component%
\begin{equation}
\left[ \tilde{\partial}_{1}^{2}-e\tilde{B}\tilde{\xi}^{2}+\left( \tilde{p}%
^{0}\right) ^{2}-m^{2}-e\tilde{B}\right] \phi _{\uparrow }\left( \tilde{x}%
^{1}\right) =0,  \label{equationphyup}
\end{equation}%
with the notation%
\begin{equation}
\tilde{\xi}=\sqrt{e\tilde{B}}\left( \tilde{x}^{1}+\frac{\tilde{p}^{2}}{e%
\tilde{B}}\right) .  \label{notation ksi}
\end{equation}%
From the standard course of quantum mechanics, it is well known that the
solution of the equation (\ref{equationphyup}), finite in the limit $\tilde{%
\xi}\rightarrow \infty $, is obtained under the condition $\left[ \left(
\tilde{p}^{0}\right) ^{2}-m^{2}\right] /e\tilde{B}=2n+2$, where $%
n=0,1,2,\ldots $. The corresponding solution is expressed in terms of the
Hermite polynomials $H_{n}(x)$ as
\begin{equation}
\phi _{\uparrow }\left( \tilde{x}^{1}\right) =\tilde{C}_{n}e^{-\frac{\tilde{%
\xi}^{2}}{2}}H_{n}\left( \tilde{\xi}\right) .  \label{solution phyup}
\end{equation}%
The lower component is obtained from (\ref{equations phys}):%
\begin{equation}
\phi _{\downarrow }\left( \tilde{x}^{1}\right) =\frac{i\tilde{C}_{n}\sqrt{e%
\tilde{B}}}{\tilde{p}_{\pm }^{0}+sm}e^{-\frac{\tilde{\xi}^{2}}{2}%
}H_{n+1}\left( \tilde{\xi}\right) ,  \label{solution phydown}
\end{equation}%
where we have introduced the notation%
\begin{equation}
\tilde{p}_{\pm }^{0}=\pm \sqrt{2\left( n+1\right) e\tilde{B}+m^{2}}.
\label{p0t}
\end{equation}

For the corresponding complete set of positive and negative energy mode
functions, we have
\begin{equation}
\tilde{\Psi}^{\pm }\left( \tilde{x}^{\mu }\right) =\left(
\begin{array}{c}
\tilde{\Psi}_{\uparrow }^{\pm }\left( \tilde{x}^{\mu }\right) \\
\tilde{\Psi}_{\downarrow }^{\pm }\left( \tilde{x}^{\mu }\right)%
\end{array}%
\right) =\tilde{C}_{n}e^{i\left( \tilde{p}^{2}\tilde{x}^{2}-\tilde{p}_{\pm
}^{0}\tilde{x}^{0}\right) }e^{-\frac{\tilde{\xi}^{2}}{2}}\left(
\begin{array}{c}
H_{n}\left( \tilde{\xi}\right) \\
\frac{i\sqrt{e\tilde{B}}}{\tilde{p}_{\pm }^{0}+sm}H_{n+1}\left( \tilde{\xi}%
\right)%
\end{array}%
\right) ,  \label{labwavefunction}
\end{equation}%
The normalization coefficient is given by
\begin{equation}
\left\vert \tilde{C}_{n}\right\vert ^{2}=\frac{\sqrt{e\tilde{B}}}{\pi ^{%
\frac{3}{2}}2^{n+1}n!}\left[ 1+\frac{2\left( n+1\right) e\tilde{B}}{\left(
\tilde{p}_{\pm }^{0}+sm\right) ^{2}}\right] ^{-1}.  \label{Cnt}
\end{equation}

By applying the inverse boost transformation to the modes (\ref%
{labwavefunction}), $\Psi ^{\pm }\left( x^{\mu }\right) =\exp \left( -\theta
\sigma _{2}/2\right) \tilde{\Psi}^{\pm }\left( \tilde{x}^{\mu }\right) $,
for the modes in our problem we get%
\begin{eqnarray}
\Psi ^{\pm }\left( x^{\mu }\right) &=&\left(
\begin{array}{c}
\Psi _{\uparrow }^{\pm }\left( x^{\mu }\right) \\
\Psi _{\downarrow }^{\pm }\left( x^{\mu }\right)%
\end{array}%
\right) =C_{n}e^{-\frac{\xi ^{2}}{2}}e^{i\left( p^{2}x^{2}-p_{\pm
}^{0}x^{0}\right) }  \notag \\
&&\times \left(
\begin{array}{c}
H_{n}\left( \xi \right) \cosh \left( \frac{\theta }{2}\right) -\frac{\sqrt{eB%
}\left( 1-\beta ^{2}\right) ^{\frac{1}{4}}}{\pm g_{n}\left( B,\beta \right)
+sm}H_{n+1}\left( \xi \right) \sinh \left( \frac{\theta }{2}\right) \\
-iH_{n}\left( \xi \right) \sinh \left( \frac{\theta }{2}\right) +\frac{i%
\sqrt{eB}\left( 1-\beta ^{2}\right) ^{\frac{1}{4}}}{\pm g_{n}\left( B,\beta
\right) +sm}H_{n+1}\left( \xi \right) \cosh \left( \frac{\theta }{2}\right)%
\end{array}%
\right) ,  \label{wave function EB}
\end{eqnarray}%
where%
\begin{equation}
p_{\pm }^{0}=\pm \sqrt{1-\beta ^{2}}g_{n}\left( B,\beta \right) -\beta p^{2},
\label{energy}
\end{equation}%
and%
\begin{equation}
\xi =\sqrt{eB\sqrt{1-\beta ^{2}}}\left( x^{1}+\frac{p^{2}}{eB}+\frac{\beta
g_{n}\left( B,\beta \right) }{\sqrt{eB}\sqrt{1-\beta ^{2}}}\right) .
\label{qsiEB}
\end{equation}%
The notation used in these expressions is defined as%
\begin{equation}
g_{n}\left( B,\beta \right) =\sqrt{2\left( n+1\right) eB\sqrt{1-\beta ^{2}}%
+m^{2}}.  \label{gn}
\end{equation}%
The normalization coefficient, written in terms of the fields $E$ and $B$,
takes the form%
\begin{equation}
\left\vert C_{n}\right\vert ^{2}=\frac{\sqrt{eB\sqrt{1-\beta ^{2}}}}{\pi ^{%
\frac{3}{2}}2^{n+1}n!}\left[ 1+\frac{2\left( n+1\right) }{\left( p_{\pm
}^{0}+sm\right) ^{2}}eB\sqrt{1-\beta ^{2}}\right] ^{-1}.  \label{Cn}
\end{equation}%
The dependence of the mode functions on the field mass is rather
complicated. It enters through the function (\ref{gn}) in the arguments of
the Hermite polynomials, in the coefficients of those polynomials in (\ref%
{wave function EB}) and in the normalization coefficient $C_{n}$. For large
values of $\xi $ the modes behave like $\xi ^{n+1}e^{-\frac{\xi ^{2}}{2}}$.

\begin{figure}[tbph]
\begin{center}
\epsfig{figure=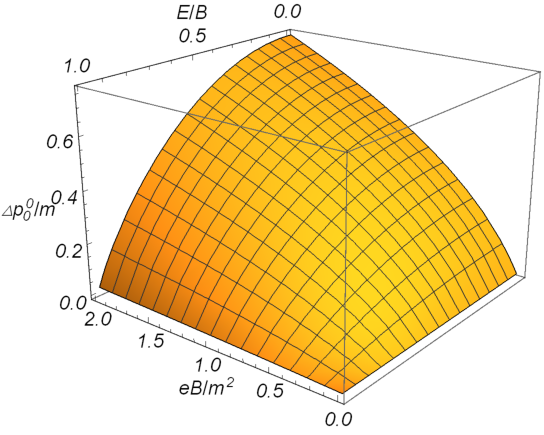,width=7.cm,height=6.cm}
\end{center}
\caption{The difference between the energies of the first excited state and
ground state versus the ratios $eB/m^{2}$ and $E/B$.}
\label{fig1}
\end{figure}

The difference of the energies between the neighboring levels is given by
the expression $\Delta p_{n}^{0}=\sqrt{1-\beta ^{2}}\left[ g_{n+1}\left(
B,\beta \right) -g_{n}\left( B,\beta \right) \right] $. In Figure \ref{fig1}
we plot this difference for $n=0$ as a function of dimensionless ratios $%
eB/m^{2}$ and $E/B$. The collapse of energy levels in the limit $%
E\rightarrow B$ is seen (see also \cite{Lukose}). In the next section, we
use the modes (\ref{wave function EB}) for the evaluation of the fermionic
condensate.

\section{Condensate for massive fermions}

\label{sec:FC}

Denoting by $\left\vert 0\right\rangle $ the vacuum state, for the fermion
condensate one has
\begin{equation}
\left\langle 0\left\vert \bar{\Psi}\Psi \right\vert 0\right\rangle
=\left\langle \bar{\Psi}\Psi \right\rangle =-\frac{1}{2}\sum_{n=0}^{\infty
}\int_{-\infty }^{+\infty }dp^{2}\sum_{\chi =-,+}\chi \bar{\Psi}^{\chi }\Psi
^{\chi },  \label{condensdefin}
\end{equation}%
where $\bar{\Psi}=\Psi ^{\dagger }\gamma ^{0}$, $\Psi ^{\pm }$ is a complete
set of positive and negative energy solutions to the Dirac equation. By
substituting the mode functions from (\ref{wave function EB}) and
integrating over $p^{2}$, we obtain

\begin{equation}
\left\langle \bar{\Psi}\Psi \right\rangle =-\frac{1}{2\pi }%
\sum_{n=0}^{\infty }\frac{smeB}{g_{n}\left( B,\beta \right) }.
\label{condenssum}
\end{equation}

This result is expressed in terms of the Hurwitz zeta function $\zeta (\nu
,u)$:
\begin{equation}
\left\langle \bar{\Psi}\Psi \right\rangle =-\frac{sm\sqrt{eB}}{2^{\frac{3}{2}%
}\pi \left( 1-\beta ^{2}\right) ^{\frac{1}{4}}}\zeta \left( \frac{1}{2}%
,u\right) ,  \label{condensHurwitz}
\end{equation}%
where

\begin{equation}
u=1+\frac{m^{2}}{2eB\sqrt{1-\beta ^{2}}}.  \label{udefinition}
\end{equation}

For zero mode, $\tilde{p}^{0}=-sm$, the expression for the wave function
reads:
\begin{equation}
\Psi _{\left( 0\right) }\left( x^{\mu }\right) =C^{(0)}\left(
\begin{array}{c}
i\sinh \left( \theta /2\right) \\
\cosh \left( \theta /2\right)%
\end{array}%
\right) e^{-\frac{\xi ^{2}}{2}}e^{i\left( p^{2}x^{2}-p^{0}x^{0}\right) },
\label{wave function zeromode}
\end{equation}%
where%
\begin{equation*}
\left\vert C^{(0)}\right\vert ^{2}=\frac{\sqrt{eB}}{2\pi ^{\frac{3}{2}}}%
\left( 1-\beta ^{2}\right) ^{\frac{1}{4}}.
\end{equation*}%
For the contribution coming from the zero modes, one gets
\begin{equation}
\left\langle \bar{\Psi}\Psi \right\rangle _{\left( 0\right) }=-\frac{eB}{%
2\pi }.  \label{condzeromode}
\end{equation}%
Combining this with (\ref{condensHurwitz}) for the total condensate we find
\begin{equation}
\left\langle \bar{\Psi}\Psi \right\rangle _{\mathrm{tot}}=-\frac{eB}{2\pi }-%
\frac{sm\sqrt{eB}}{2^{\frac{3}{2}}\pi \left( 1-\beta ^{2}\right) ^{\frac{1}{4%
}}}\zeta \left( \frac{1}{2},u\right) .  \label{condtotHurwitz}
\end{equation}

For unique renormalized vacuum expectation value additional renormalization
conditions should be imposed. As such conditions, we require zero vacuum
expectation values in the absence of external electromagnetic fields. This
procedure is reduced to the subtraction from the expression given above its
limiting value when $B,$ $E\rightarrow 0$. From (\ref{condtotHurwitz}), by
taking into account that
\begin{equation}
\zeta (s,a)\approx \frac{a^{1-s}}{s-1},  \label{Zetas}
\end{equation}%
for $u\rightarrow \infty $ (see \cite{Olver}), one obtains
\begin{equation}
\lim_{B,E\rightarrow 0}\left\langle \bar{\Psi}\Psi \right\rangle _{\mathrm{%
tot}}=\frac{sm^{2}}{2\pi }.  \label{condlim}
\end{equation}%
As a result, the renormalized fermion condensate is presented in the form
\begin{equation}
\left\langle \bar{\Psi}\Psi \right\rangle _{\mathrm{ren}}=-\frac{eB}{2\pi }-%
\frac{sm\sqrt{eB}}{2^{\frac{3}{2}}\pi \left( 1-\beta ^{2}\right) ^{\frac{1}{4%
}}}\zeta \left( \frac{1}{2},u\right) -\frac{sm^{2}}{2\pi },
\label{condtotrenorm}
\end{equation}%
where $u$ is defined in (\ref{udefinition}). Note that the additional
renormalization condition $\lim_{B,E\rightarrow 0}\left\langle \bar{\Psi}%
\Psi \right\rangle _{\mathrm{ren}}=0$ is similar to that used in the
renormalization of the vacuum expectation values in the Casimir effect (see,
for instance, \cite{Bordag}). It is reduced to the requirement that the
vacuum expectation values should vanish in the Minkowski spacetime in the
absence of additional boundaries (in the context of the Casimir effect) and
external fields (in the present context). Further discussion on the finite
renormalization terms in problems with smooth background fields can be found
in \cite{Bordag}. Note that a similar renormalization condition $%
\lim_{B\rightarrow 0}\left\langle \bar{\Psi}\Psi \right\rangle _{\mathrm{ren}%
}=0$ has been used in \cite{Ditt97} for the fermion condensate in the
problem with zero electric field.

In the absence of the electric field the formula (\ref{condtotrenorm}) is
reduced to%
\begin{equation}
\left\langle \bar{\Psi}\Psi \right\rangle _{\mathrm{ren}}|_{E=0}=-\frac{eB}{%
2\pi }-\frac{sm\sqrt{eB}}{2^{\frac{3}{2}}\pi }\zeta \left( \frac{1}{2},1+%
\frac{m^{2}}{2eB}\right) -\frac{sm^{2}}{2\pi }.  \label{FCE0}
\end{equation}%
The first two terms on the right-hand side with $s=1$ differ from the
corresponding terms in \cite{Ditt97} by an additional coefficient of 1/2.
The reason for that difference is related to the fact that in \cite{Ditt97}
the condensate is given for 4-component spinors. The last term in (\ref{FCE0}%
) differs by the sign (in addition to the factor 1/2 explained above) from
the corresponding term in \cite{Ditt97} and the result in that reference
does not obey the renormalization condition we have imposed on the fermion
condensate in the absence of external fields. For the behavior of the
condensate in the limit $E\rightarrow B$ ($\beta \rightarrow 1$), again, we
use the asymptotic expression (\ref{Zetas}). For the leading term, this
gives
\begin{equation}
\left\langle \bar{\Psi}\Psi \right\rangle _{\mathrm{ren}}\approx \frac{sm^{2}%
}{2\pi \sqrt{1-\beta ^{2}}}.  \label{condassym}
\end{equation}%
In a similar way, for the asymptotic in the range $m^{2}\gg eB$ we get $%
\left\langle \bar{\Psi}\Psi \right\rangle \approx sm^{2}(1/\sqrt{1-\beta ^{2}%
}-1)/(2\pi )$. In the case of $m=0$, there is no fermionic condensate.

\begin{figure}[tbph]
\begin{center}
\epsfig{figure=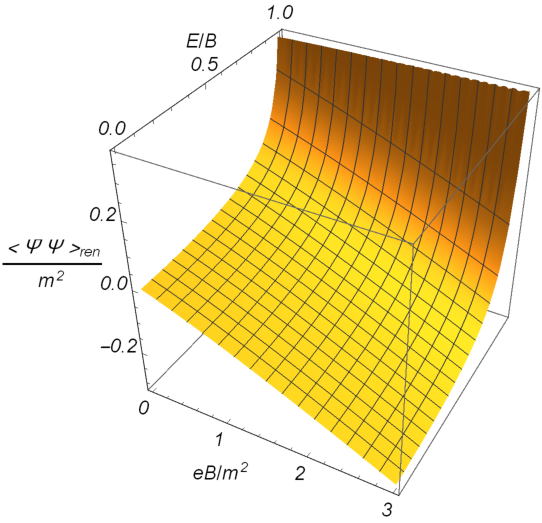,width=7.cm,height=6.cm}
\end{center}
\caption{The renormalized fermion condensate in units of $m^{2}$, $%
\left\langle \overline{\Psi }\Psi \right\rangle _{\mathrm{ren}}/m^{2}$,
versus $eB/m^{2}$ and $E/B$ for the field with $s=1$.}
\label{fig2}
\end{figure}

We note that the ratio $\left\langle \bar{\Psi}\Psi \right\rangle _{\mathrm{%
ren}}/m^{2}$ depends on $m$, $B$, and $E$ through two dimensionless
combinations $eB/m^{2}$ and $E/B$. In Figure \ref{fig2} we have plotted the
renormalized fermion condensate (in units of $m^{2}$) for the case $s=1$ as
a function of $eB/m^{2}$ and $E/B$. As seen, depending on the values of the
fields the fermion condensate can be either positive or negative. In the
absence of the electric field, the condensate is negative. A sufficiently
strong electric field leads to the change of the sign. As we can see in
Figure \ref{fig2}, the condensate changes the sign (for a similar situation,
in the fermionic condensate in de Sitter spacetime, see, e.g., \cite{Sah}).
The fermionic condensate is an important physical characteristic in
interacting field theories. An example is the Nambu-Jona-Lasinio model with
the self-interaction of the type $g(\bar{\Psi}\Psi )^{2}$, where $g$ is the
coupling constant. Another type of fermionic model interacting with a scalar
field $\varphi (x)$ through the potential $\varphi ^{2}\bar{\Psi}\Psi $ has
been discussed in \cite{Sah}. The generation of the fermionic condensate
induces an additional mass term in the effective field equation for the
field $\varphi (x)$. The mass squared of the field is shifted by the
contribution proportional to $\left\langle \bar{\Psi}\Psi \right\rangle _{%
\mathrm{ren}}$. The change in the sign of condensate may change the sign of
the effective mass squared with a possible phase transition (see also \cite%
{Sah}).

\section{VEV of the energy-momentum tensor}

\label{sec:EMT}

In this Section, we consider another important characteristic of the
fermionic vacuum, namely, the VEV of the energy-momentum tensor. The
corresponding operator is given as
\begin{equation}
T_{\mu \nu }=\frac{1}{2}i\left[ \bar{\Psi}\gamma _{(\mu }D_{\nu )}\Psi
-\left( D_{(\mu }\bar{\Psi}\right) \gamma _{\nu )}\Psi \right] ,
\label{energymomentum tensor}
\end{equation}%
where
\begin{equation}
D_{\mu }=\partial _{\mu }+ieA_{\mu },  \label{Demu}
\end{equation}%
and the brackets in the indices mean symmetrization over the included
indices.

The corresponding VEV is written in the form of the mode sum%
\begin{equation}
\left\langle T_{\mu \nu }\right\rangle =-\frac{i}{4}\sum_{\sigma }\sum_{\chi
=-,+}\chi \left[ \bar{\Psi}_{\sigma }^{\chi }\gamma _{(\mu }D_{\nu )}\Psi
_{\sigma }^{\chi }-\left( D_{(\mu }\bar{\Psi}_{\sigma }^{\chi }\right)
\gamma _{\nu )}\Psi _{\sigma }^{\chi }\right] .  \label{Tmunudefinition}
\end{equation}%
By using the mode functions given above it can be seen that the off-diagonal
components of the energy-momentum tensor vanish. Therefore, let us consider
the diagonal components of the energy-momentum tensor.

For the VEV of the energy density we get
\begin{equation}
\left\langle T_{00}\right\rangle =-\frac{1}{2}\sum_{\sigma }\int_{-\infty
}^{+\infty }dp^{2}\sum_{\chi =-,+}\chi \left( p^{0\left( \chi \right)
}-eA_{0}\right) \Psi _{\sigma }^{\chi \dagger }\Psi _{\sigma }^{\chi }.
\label{T00 definition}
\end{equation}%
Substituting the mode functions and using the orthogonality relation for the
Hermite polynomials, we obtain the following expression for the VEV of the
energy density:
\begin{equation}
\left\langle T_{00}\right\rangle =-\frac{eB}{2\pi }\sum_{n=0}^{\infty
}g_{n}\left( B,\beta \right) \left( 1+\frac{\beta }{\sqrt{1-\beta ^{2}}}%
\frac{\left( n+1\right) eB}{2\left( n+1\right) eB\sqrt{1-\beta ^{2}}+m^{2}}%
\right) .  \label{T00}
\end{equation}%
For the contribution of the zero mode (\ref{wave function zeromode}) to the
energy density we get
\begin{equation}
\left\langle T_{00}\right\rangle _{(0)}=-\frac{eB}{2\pi }\frac{m}{1-\beta
^{2}}.  \label{T00zeromode}
\end{equation}%
The total VEV of the energy density is expressed as
\begin{equation}
\left\langle T_{00}\right\rangle _{\mathrm{tot}}=-\frac{eB}{2\pi }\frac{m}{%
1-\beta ^{2}}-\frac{eB}{2\pi }\sum_{n=0}^{+\infty }g_{n}\left( B,\beta
\right) \left[ 1+\frac{\beta }{\sqrt{1-\beta ^{2}}}\frac{\left( n+1\right) eB%
}{2\left( n+1\right) eB\sqrt{1-\beta ^{2}}+m^{2}}\right] .  \label{T00tot}
\end{equation}%
This result is expressed in terms of the Hurwitz function
\begin{eqnarray}
\left\langle T_{00}\right\rangle _{\mathrm{tot}} &=&\frac{eB}{2\pi }\frac{m}{%
1-\beta ^{2}}\left\{ -1+\frac{\left( 1-\beta ^{2}\right) ^{-\frac{1}{4}}}{%
\sqrt{2}}\left[ \frac{m\beta }{2\sqrt{eB}}\zeta \left( \frac{1}{2},u\right)
\right. \right.  \notag \\
&&\left. \left. -\left( 2+\frac{\beta }{1-\beta ^{2}}\right) \left( 1-\beta
^{2}\right) ^{\frac{3}{2}}\frac{\sqrt{eB}}{m}\zeta \left( -\frac{1}{2}%
,u\right) \right] \right\} .  \label{T00totHurwitz}
\end{eqnarray}%
The vacuum energy density is the same for the fields with $s=1$ and $s=-1$.

In the case of the absence of electric and magnetic fields we have
\begin{equation}
\lim_{B,E\rightarrow 0}\left\langle T_{00}\right\rangle _{\mathrm{tot}}=%
\frac{m^{3}}{6\pi }.  \label{T00lim}
\end{equation}%
Similar to the case of the fermion condensate we impose a renormalization
condition that corresponds to the zero energy density in the limit $B,$ $%
E\rightarrow 0$. By taking into account (\ref{T00lim}), for the renormalized
energy density one obtains
\begin{eqnarray}
\left\langle T_{00}\right\rangle _{\mathrm{ren}} &=&\frac{eB}{2\pi }\frac{m}{%
1-\beta ^{2}}\left\{ -1+\frac{\left( 1-\beta ^{2}\right) ^{-\frac{1}{4}}}{%
\sqrt{2}}\left[ \frac{m\beta }{2\sqrt{eB}}\zeta \left( \frac{1}{2},u\right)
\right. \right.  \notag \\
&&\left. \left. -\left( 2+\frac{\beta }{1-\beta ^{2}}\right) \left( 1-\beta
^{2}\right) ^{\frac{3}{2}}\frac{\sqrt{eB}}{m}\zeta \left( -\frac{1}{2}%
,u\right) \right] \right\} -\frac{m^{3}}{6\pi }.  \label{T00totHurvitzassym}
\end{eqnarray}%
In the special case of the zero electric field, this expression is
simplified to
\begin{equation}
\left\langle T_{00}\right\rangle _{\mathrm{ren}}=-\frac{eB}{2\pi }\left[ m+%
\sqrt{2eB}\zeta \left( -\frac{1}{2},1+\frac{m^{2}}{2eB}\right) \right] -%
\frac{m^{3}}{6\pi }.  \label{T00E0}
\end{equation}%
In the limit when the electric field tends to its critical value, $%
E\rightarrow B$, one has $\beta \rightarrow 1$ and by using the asymptotic (%
\ref{Zetas}) for the leading term in the expansion of the energy density we
find
\begin{equation}
\left\langle T_{00}\right\rangle _{\mathrm{ren}}\approx -\frac{m^{3}}{6\pi
\left( 1-\beta ^{2}\right) ^{\frac{3}{2}}}.  \label{T00totassym}
\end{equation}

\begin{figure}[tbph]
\begin{center}
\epsfig{figure=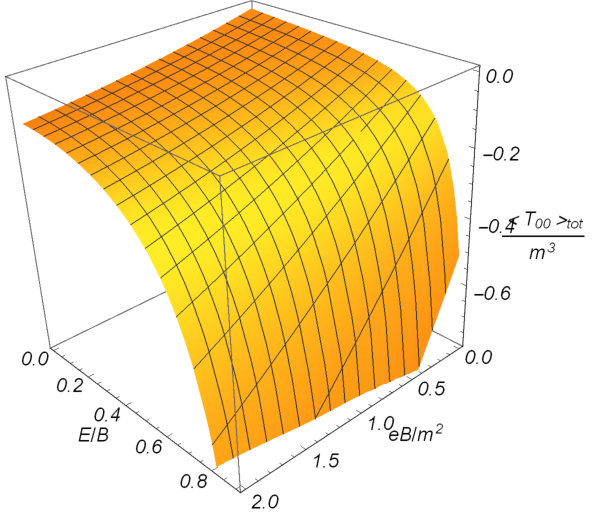,width=7.cm,height=6.cm}
\end{center}
\caption{The renormalized VEV of the energy density in units of $m^{3}$, $%
\left\langle T_{00}\right\rangle _{\mathrm{ren}}/m^{3}$, as a function of $%
eB/m^{2}$ and $E/B$.}
\label{fig3}
\end{figure}

The ratio $\left\langle T_{00}\right\rangle _{\mathrm{ren}}/m^{3}$ is a
function of the ratios $eB/m^{2}$ and $E/B$ (the same is the case for the
spatial components). In Figure \ref{fig3} the dependence of the energy
density (in units of $m^{3}$) is plotted on the combinations $eB/m^{2}$ and $%
E/B$. As it is seen from the graph, the energy density is negative. The
emergence of negative energy density in a quantum field theory is very
legitimate. The renormalized energy density of a quantized field is obtained
by the subtraction of divergent contribution and, in general, is not
positive (see, e.g., \cite{Ford}). Well known example is the energy density
for the electromagnetic vacuum in the region between two parallel conducting
plates (the Casimir effect) \cite{Bordag}.

Now we consider the component
\begin{equation}
\left\langle T_{11}\right\rangle =-\frac{i}{4}\sum_{\sigma }\sum_{\chi
=-,+}\chi \left[ \Psi _{\sigma }^{\chi \dagger }\gamma ^{0}\gamma
_{1}\partial _{1}\Psi _{\sigma }^{\chi }-\left( \partial _{1}\Psi _{\sigma
}^{\chi \dagger }\right) \gamma ^{0}\gamma _{1}\Psi _{\sigma }^{\chi }\right]
.  \label{T11defin}
\end{equation}%
With the mode functions from (\ref{wave function EB}) this expression is
transformed to

\begin{equation}
\left\langle T_{11}\right\rangle =-\frac{\left( eB\right) ^{2}}{2\pi }\sqrt{%
1-\beta ^{2}}\sum_{n=0}^{\infty }\frac{n+1}{g_{n}\left( B,\beta \right) }.
\label{T11}
\end{equation}%
The contribution to the component $\left\langle T_{11}\right\rangle $ coming
from the zero mode vanishes. As before, we write the corresponding
expression in terms of the Hurwitz function:
\begin{equation}
\left\langle T_{11}\right\rangle =-\frac{\left( eB\right) ^{\frac{3}{2}}}{2^{%
\frac{3}{2}}\pi }\left( 1-\beta ^{2}\right) ^{\frac{1}{4}}\left[ \zeta
\left( -\frac{1}{2},u\right) -\frac{m^{2}}{2eB\sqrt{1-\beta ^{2}}}\zeta
\left( \frac{1}{2},u\right) \right] .  \label{T11Hurwitz}
\end{equation}%
Again, in the case of the absence of electric and magnetic fields we have
the following limiting value:
\begin{equation}
\lim_{B,E\rightarrow 0}\left\langle T_{11}\right\rangle =-\frac{m^{3}}{6\pi }%
.  \label{T11lim}
\end{equation}%
The renormalization condition described above leads to the renormalized VEV
\begin{equation}
\left\langle T_{11}\right\rangle _{\mathrm{ren}}=-\frac{\left( eB\right) ^{%
\frac{3}{2}}}{2^{\frac{3}{2}}\pi }\left( 1-\beta ^{2}\right) ^{\frac{1}{4}}%
\left[ \zeta \left( -\frac{1}{2},u\right) -\frac{m^{2}}{2eB\sqrt{1-\beta ^{2}%
}}\zeta \left( \frac{1}{2},u\right) \right] +\frac{m^{3}}{6\pi }.
\label{T11renorm}
\end{equation}

In the problem with zero electric field the formula (\ref{T11renorm}) is
rewritten as
\begin{equation}
\left\langle T_{11}\right\rangle _{\mathrm{ren}}=-\frac{\left( eB\right) ^{%
\frac{3}{2}}}{2^{\frac{3}{2}}\pi }\left[ \zeta \left( -\frac{1}{2},1+\frac{%
m^{2}}{2eB}\right) -\frac{m^{2}}{2eB}\zeta \left( \frac{1}{2},1+\frac{m^{2}}{%
2eB}\right) \right] +\frac{m^{3}}{6\pi }.  \label{T11E0}
\end{equation}%
For the asymptotic behavior of the VEV of the component $\left\langle
T_{11}\right\rangle $ in the limit $\beta \rightarrow 1$ one obtains
\begin{equation}
\left\langle T_{11}\right\rangle _{\mathrm{ren}}\approx -\frac{m^{3}}{6\pi
\sqrt{1-\beta ^{2}}}.  \label{T11assymp}
\end{equation}

It remains to consider the 22-component
\begin{equation}
\left\langle T_{22}\right\rangle =\frac{1}{2}\sum_{\sigma }\sum_{\chi
=-,+}\chi \left( p^{2}+eBx^{1}\right) \Psi _{\sigma }^{\chi \dagger }\gamma
^{0}\gamma _{2}\Psi _{\sigma }^{\chi }.  \label{T22defin}
\end{equation}%
With the mode functions given above, this component is expressed as%
\begin{equation}
\left\langle T_{22}\right\rangle =-\frac{\left( eB\right) ^{2}}{2\pi \sqrt{%
1-\beta ^{2}}}\sum_{n=0}^{\infty }\frac{n+1}{g_{n}\left( B,\beta \right) }.
\label{T22b}
\end{equation}%
Written in terms of the Hurwitz function, one gets
\begin{equation}
\left\langle T_{22}\right\rangle =-\frac{\left( eB\right) ^{\frac{3}{2}}}{2^{%
\frac{3}{2}}\pi }\frac{1}{\left( 1-\beta ^{2}\right) ^{\frac{3}{4}}}\left[
\zeta \left( -\frac{1}{2},u\right) -\frac{m^{2}}{2eB\sqrt{1-\beta ^{2}}}%
\zeta \left( \frac{1}{2},u\right) \right] .  \label{T22Hurwitz}
\end{equation}%
Taking into account that%
\begin{equation}
\lim_{B,E\rightarrow 0}\left\langle T_{22}\right\rangle =-\frac{m^{3}}{6\pi }%
,  \label{T22lim}
\end{equation}%
for the renormalized VEV we obtain
\begin{equation}
\left\langle T_{22}\right\rangle _{\mathrm{ren}}=-\frac{\left( eB\right) ^{%
\frac{3}{2}}}{2^{\frac{3}{2}}\pi }\frac{1}{\left( 1-\beta ^{2}\right) ^{%
\frac{3}{4}}}\left[ \zeta \left( -\frac{1}{2},u\right) -\frac{m^{2}}{2eB%
\sqrt{1-\beta ^{2}}}\zeta \left( \frac{1}{2},u\right) \right] +\frac{m^{3}}{%
6\pi }.  \label{T22renorm}
\end{equation}%
In the absence of electric field $\left\langle T_{22}\right\rangle _{\mathrm{%
ren}}=\left\langle T_{11}\right\rangle _{\mathrm{ren}}$ which is a direct
consequence of the symmetry and calculating the trace of the energy-momentum
tensor from formulas (\ref{T00E0}), (\ref{T11E0}), (\ref{T22renorm}) and
passing to the $E\rightarrow 0$ limit, we will get the result given in \cite%
{Ditt97} (the difference in the sign is due to the difference in the
signature). Note that for $1-\beta ^{2}\ll 1$ one has $\left\langle
T_{22}\right\rangle _{\mathrm{ren}}\approx \left\langle T_{00}\right\rangle
_{\mathrm{ren}}$ and $|\left\langle T_{11}\right\rangle _{\mathrm{ren}}|\ll
|\left\langle T_{22}\right\rangle _{\mathrm{ren}}|$. We note once again that
the renormalization condition $\lim_{B,E\rightarrow 0}\left\langle T_{\mu
\nu }\right\rangle _{\mathrm{ren}}=0$ has been imposed that corresponds to
the zero expectation value of the energy-momentum tensor in Minkowski
spacetime in the absence of external fields. This condition is required also
by the Einstein equations for the Minkowskian geometry.

\begin{figure}[tbph]
\begin{center}
\begin{tabular}{cc}
\epsfig{figure=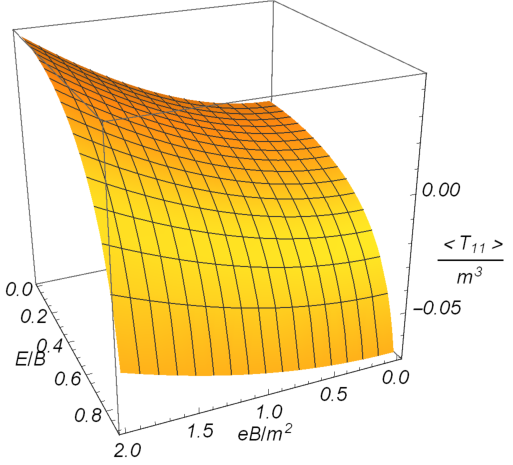,width=7.cm,height=6.cm} & \quad %
\epsfig{figure=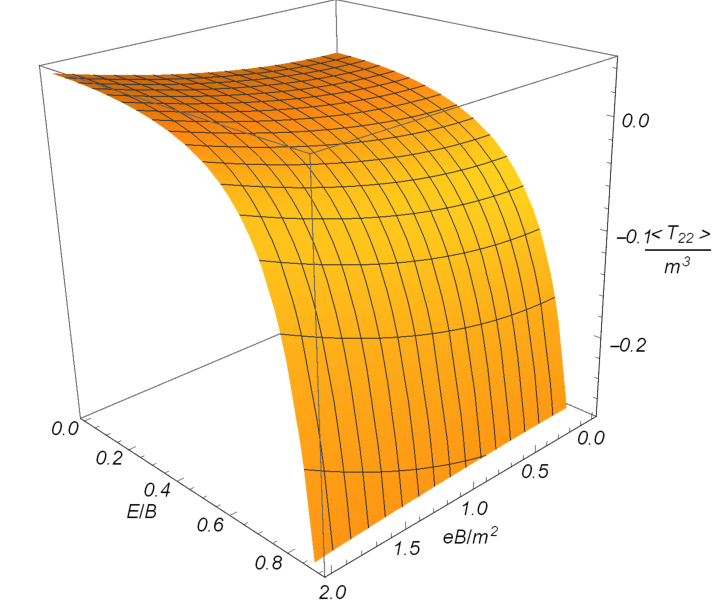,width=7.cm,height=6.cm}%
\end{tabular}%
\end{center}
\caption{The dependence of the spatial components of the vacuum
energy-momentum tensor (in units of $m^{3}$) on the ratios $eB/m^{2}$ and $%
E/B$. }
\label{fig4}
\end{figure}

The Figure \ref{fig4} displays the components $\left\langle
T_{11}\right\rangle _{\mathrm{ren}}$ (left panel) and $\left\langle
T_{22}\right\rangle _{\mathrm{ren}}$ (right panel) as functions of
dimensionless combinations $eB/m^{2}$ and $E/B$. Both the components are
positive for $E=0$ and with increasing $E$ they change the sign.

General formulas are being simplified for a massless field. Taking the limit
$m\rightarrow 0$ in (\ref{T00totHurvitzassym}),(\ref{T11renorm}),(\ref%
{T22renorm}), we get%
\begin{equation}
\left\langle T_{\mu \nu }\right\rangle _{\mathrm{ren}}=-\frac{\left(
eB\right) ^{\frac{3}{2}}}{2^{\frac{3}{2}}\pi }\zeta \left( -\frac{1}{2}%
\right) \mathrm{diag}\left[ \left( 1-\beta ^{2}\right) ^{\frac{1}{4}}\left(
2+\frac{\beta }{1-\beta ^{2}}\right) ,1,\left( 1-\beta ^{2}\right) ^{-\frac{3%
}{4}}\right] ,  \label{equationmzero}
\end{equation}%
where $\zeta \left( x\right) $ is Riemann zeta function with $\zeta
(-1/2)\approx -0.208$.

In a graphene sheet, the long-wavelength description of electronic states
can be formulated in terms of the Dirac theory of spinors in
(2+1)-dimensional space-time, where the Fermi velocity plays the role of the
speed of light. In the simplest setup, the corresponding quasiparticles are
massless. In this special case, the condensate vanishes and the
energy-momentum tensor is reduced to (\ref{equationmzero}). However, it
should be noted that the energy gap, corresponding to the mass term in the
effective Dirac equation, can be generated by a number of mechanisms (see,
for instance, \cite{Gusy2007, Cham00,Seme84,Giov07} and references therein).
The latter include the deformations of bonds in the lattice of graphene and
the symmetry breaking between two triangular sublattices by staggered
on-site energy. The mass term is also generated by the interaction of a
graphene monolayer with a substrate that breaks the sublattice symmetry. The
expression for the fermion condensate and of the energy-momentum tensor in
the corresponding massive model are obtained from the formulae given above
by the replacements $\beta \rightarrow E/(v_{F}B)$ and $m\rightarrow $ $%
mv_{F}$.

As it has been shown the vacuum energy density is negative. In the
literature, several constraints on negative energy in the form of
inequalities were discussed (see, e.g., \cite{Ford,Fewst05}). It will be
interesting to consider the validity of those constraints in the problem
under consideration. The constraints for the Dirac field have been studied
in \cite{Voll00} and \cite{Fews03,Yu03} in 2- and 4-dimensional flat
spacetimes in the absence of external fields. The case of the general number
of spatial dimensions is analyzed in \cite{Xing06}. The quantum weak energy
inequalities for the Dirac field in the background of four-dimensional
globally hyperbolic spacetime were established in \cite{Fews02}. The
corresponding constraints (quantum energy inequalities (QEI)) contain smooth
sampling functions. The consideration of QEI and their dependence on the
form of the sampling function in the problem under consideration is
additionally complicated by the presence of an external electromagnetic
field and requires a separate investigation.

\section{Conclusion}

\label{sec:Conc}

In this paper, we have investigated the massive fermionic quantum field
localized on a plane in external constant and homogeneous electric and
magnetic fields. The magnetic field is perpendicular to the plane and the
electric field is parallel to the plane. For the evaluation of the fermionic
vacuum characteristics, we have used the summation over the corresponding
mode functions. Those functions are presented in section 2. The related
energy spectrum and its features have been discussed in \cite{Lukose}. Here
we investigate the fermion condensate and the VEV of the energy-momentum
tensor. The renormalization procedure is based on the zeta function
technique. We are interested in the effects induced by external electric and
magnetic fields and an additional renormalization condition is imposed that
requires zero expectation values in the absence of external fields. The
renormalized fermion condensate is negative for zero electric field and
becomes positive for the values of the electric field close to the critical
value $E\rightarrow B$ ($\beta \rightarrow 1$). The vacuum energy density is
always negative, whereas the vacuum stresses are positive in the region $%
E\ll B$ and negative for $1-(E/B)^{2}\ll 1$. For the values of the electric
field near the critical value, $E\rightarrow B$ one has $|\left\langle
T_{11}\right\rangle _{\mathrm{ren}}|\ll |\left\langle T_{22}\right\rangle _{%
\mathrm{ren}}|$ and $\left\langle T_{22}\right\rangle _{\mathrm{ren}}\approx
\left\langle T_{00}\right\rangle _{\mathrm{ren}}$.

The analysis showed that in the presence of crossed uniform electric and
magnetic fields, we have the following phenomena: the contraction and
eventual collapse of the Landau level (see \cite{Lukose}), the changing sign
of the condensate, and the negative energy density, which can be verified
experimentally.

Another class of effects on the vacuum of (2+1)-dimensional fermions induced
by magnetic fields have been discussed in \cite{Bell09,Bell10} for
cylindrical and toroidal topologies of the background space. In those
references, the influence of the magnetic flux threading the compact
dimensions is studied and the effect has a topological nature (of the
Aharonov-Bohm type). Similar effects for conical geometry with edges were
considered in \cite{Beze10}-\cite{Bell20}. In these geometries, the local
characteristics of the vacuum state are periodic functions of the magnetic
flux with the period equal to the quantum of flux.

\end{document}